\documentclass[conference,compsoc]{IEEEtran}
\AtBeginDocument{%
  \providecommand\BibTeX{{%
    \normalfont B\kern-0.5em{\scshape i\kern-0.25em b}\kern-0.8em\TeX}}}

\usepackage{tikz}
\usepackage{amsmath}
\usepackage{multirow}
\usepackage{filecontents}
\usepackage{booktabs}
\usepackage{graphicx}
\usepackage{units}
\usepackage{blindtext}
\usepackage{xcolor,xspace}
\usepackage{amsmath}
\usepackage{paralist}
\usepackage{mdframed}
\usepackage{pict2e}
\usepackage{mdframed}
\usepackage{listings}
\usepackage{xcolor}
\usepackage{threeparttable}
\usepackage{array}
\usepackage{pifont}
\usepackage{balance}
\usepackage{verbatim}
\usepackage{array}
\usepackage{booktabs}
\usepackage{float}
\usepackage{listings}
\usepackage{hyperref}
\usepackage{balance}
\def\BibTeX{{\rm B\kern-.05em{\sc i\kern-.025em b}\kern-.08em
    T\kern-.1667em\lower.7ex\hbox{E}\kern-.125em X}}

\usepackage{nccmath}
\usepackage{soul}
\usepackage{array}
\usepackage{tabularx}

\definecolor{littlegreen}{RGB}{193,236,193}
\definecolor{littlegrey}{RGB}{128,128,128}
\definecolor{altgreen}{RGB}{0,150,0}

\usetikzlibrary{arrows}
\usetikzlibrary{shapes}
\newcommand*\circled[3]{\tikz[baseline=(char.base)]{
            \node[scale=.9,shape=circle,draw,inner sep=1pt,fill=#2,minimum size=.1cm,text=#3] (char) {#1};}}

\usepackage{color,colortbl}
\usepackage{enumitem}

\newlist{myitemize}{itemize}{1}
\setlist[myitemize]{
    label=\textbullet,
    align=left,
    leftmargin=*,
    nosep,
}

\newlist{myenumerate}{enumerate}{1}
\setlist[myenumerate]{
  label=\arabic*),
  align=left,
  leftmargin=*,
  nosep,
}

\usepackage{graphicx}
\usepackage{subcaption}

\newcommand{\ignore}[1]{}

\newcommand{\longacro}{\underline{T}EE-based \underline{R}untime \underline{A}uditing for \underline{C}ommodity \underline{E}mbedded \underline{S}ystems\xspace}
\newcommand{\acro}{\emph{TRACES}\xspace}
\newcommand{\acroplain}{TRACES\xspace}

\newcommand{\app}{{\ensuremath{\sf{App}}}\xspace}

\newcommand{\dev}{{\ensuremath{\sf{ Prv}}}\xspace}

\newcommand{\prv}{{\ensuremath{\sf{Prv}}}\xspace}
\newcommand{\vrf}{{\ensuremath{\sf{Vrf}}}\xspace}
\newcommand{\RA}{{\ensuremath{\sf{ RA}}}\xspace}
\newcommand{\PoX}{{\ensuremath{\sf{ PoX}}}\xspace}
\newcommand{\SAU}{{\ensuremath{\sf{ SAU}}}\xspace}
\newcommand{\CFA}{{\ensuremath{\sf{ CFA}}}\xspace}

\newcommand{\CFG}{{\ensuremath{\sf{ CFG}}}\xspace}

\newcommand{\adv}{{\ensuremath{\sf{ Adv}}}\xspace}
\newcommand{\chal}{{\ensuremath{\sf{ Chal}}}\xspace}

\newcommand{\CFlog}{{\ensuremath{\sf{ CF_{Log}}}}\xspace}
\newcommand{\cflog}{{\ensuremath{\sf{ CF_{Log}}}}\xspace}

\definecolor{codegreen}{rgb}{0,0.6,0}
\definecolor{codegray}{rgb}{0.5,0.5,0.5}
\definecolor{codepurple}{rgb}{0.58,0,0.82}
\definecolor{backcolour}{rgb}{0.95,0.95,0.92}

\lstdefinestyle{mystyle}{
    commentstyle=\color{codegreen},
    keywordstyle=\color{blue},
    numberstyle=\tiny\color{codegray},
    stringstyle=\color{codepurple},
    basicstyle=\fontsize{8}{8}\ttfamily,
    frame=single,
    breakatwhitespace=false,         
    breaklines=true,                 
    captionpos=b,                    
    keepspaces=true,                 
    numbers=none,                    
    numbersep=4pt,                  
    showspaces=false,                
    showstringspaces=false,
    showtabs=false,                  
    tabsize=1,
    xleftmargin=0.03\columnwidth,
    xrightmargin=0.03\columnwidth,
}

\newcounter{protocol}

\newenvironment{protocol}[1]
{
  \par\addvspace{\topsep}
  \noindent
  \tabularx{\linewidth}{@{} X @{}}
  \hline
  \refstepcounter{protocol}\textbf{Protocol \theprotocol} #1 \\
  \hline
}
{
 \endtabularx
 \par\addvspace{\topsep}
}

\newcommand{\sbline}{\\[.5\normalbaselineskip]}

\begin{document}


\title{\acroplain: TEE-based Runtime Auditing for Commodity Embedded Systems}
\date{}

\author{
\IEEEauthorblockN{Adam Caulfield\IEEEauthorrefmark{1}, Antonio Joia Neto\IEEEauthorrefmark{1}, Norrathep Rattanavipanon\IEEEauthorrefmark{2} and Ivan De Oliveira Nunes\IEEEauthorrefmark{1}}
\IEEEauthorblockA{\IEEEauthorrefmark{1}Rochester Institute of Technology, USA; \IEEEauthorrefmark{2}Prince of Songkla University, Thailand}
}

\maketitle

\begin{abstract}
Control Flow Attestation (\CFA) offers a means to detect control flow hijacking attacks on remote devices, enabling verification of their runtime trustworthiness.
\CFA generates a trace (\CFlog) containing the destination of all branching instructions executed. This allows a remote Verifier (\vrf) to inspect the execution control flow on a potentially compromised Prover (\prv) before trusting that a value/action was correctly produced/performed by \prv.
However, while \CFA can be used to detect runtime compromises, it cannot guarantee the eventual delivery of the execution evidence (\CFlog) to \vrf. In turn, a compromised \prv may refuse to send \CFlog to \vrf, preventing its analysis to determine the exploit's root cause and appropriate remediation actions.

In this work, we propose \acro: \longacro. \acro guarantees reliable delivery of periodic runtime reports even when \prv is compromised. This enables secure runtime auditing in addition to best-effort delivery of evidence in \CFA. \acro also supports a guaranteed remediation phase, triggered upon compromise detection to ensure that identified runtime vulnerabilities can be reliably patched. To the best of our knowledge, \acro is the first system to provide this functionality on commodity devices (i.e., without requiring custom hardware modifications). To that end, \acro leverages support from the ARM TrustZone-M Trusted Execution Environment (TEE). To assess practicality, we implement and evaluate a fully functional (open-source) prototype of \acro atop the commodity ARM Cortex-M33 micro-controller unit.
\end{abstract}

\section{Introduction}\label{sec:intro}

\begin{figure}[!b]
\vspace{-1.5em}
\hrulefill \\
\textbf{To appear: \textit{The 40th Annual Computer Security Applications Conference (ACSAC'24)}}
\end{figure}

Embedded devices have become ubiquitous and play critical roles within larger systems. These devices are typically implemented using resource-constrained micro-controller units (MCUs) that prioritize energy and space efficiency, as well as low cost. Due to these budgetary limitations, they lack security mechanisms commonly found in higher-end application computers, including Memory Management Units (MMUs), strong privilege separation, and inter-process isolation. Consequently, embedded devices tend to be more vulnerable to a wide range of attacks~\cite{deogirikar2017security,stuxnet,giraldo2016integrity, nafees2023smart, kayan2022cybersecurity}.

Remote Attestation (\RA)~\cite{verif_integrity_mini_survey, smart, vrased, tytan, trustlite, simple, hydra, rata, DAA, Sancus17, scraps, cra, pistis, reserve, sacha, francillon2014minimalist, brasser2016remote} has been proposed as an inexpensive means to remotely verify the software integrity of MCUs. \RA is a challenge-response protocol wherein a trusted  Verifier (\vrf) issues a cryptographic challenge and requests a timely response from a potentially compromised remote Prover device (\prv). In \RA, a root of trust within \prv is responsible for producing evidence of \prv's software state by computing an authenticated integrity check (e.g. a MAC or signature) over the current snapshot of \prv's program memory and the received challenge. By examining the produced response message, \vrf can determine if \prv's software has been illegally modified.

Although classic \RA can detect illegal program memory modifications, it cannot detect runtime attacks that do not modify code~\cite{cflat}. For instance, an adversary (\adv) could exploit a vulnerability (e.g. a buffer overflow) to hijack a program's control flow without modifying its code~\cite{war_in_mem}. Consequently, \adv could execute a malicious code sequence (e.g., Jump-/Return-Oriented Programming -- JOP/ROP -- attacks~\cite{rop}) and remain oblivious to \RA.

%

Control Flow Attestation (CFA)~\cite{sok_cfa_cfi} was introduced to augment classic \RA evidence to include an authenticated trace (denoted \CFlog) of the attested software's most recent execution. \CFlog contains all control flow transfers executed (due to branching instructions such as \texttt{jumps}, \texttt{returns}, \texttt{calls}, etc.), allowing \vrf to learn the exact path followed (see Sec.~\ref{subsec:background_cfa} for more details on \CFA).
\CFlog is usually generated by instrumenting each branching instruction in the attested binary~\cite{cflat,oat,tinycfa,iscflat} or by using custom hardware to detect and store the source/destination of branching instructions~\cite{lofat, dessouky2018litehax, zeitouni2017atrium,acfa}. As custom hardware is not yet present on commodity devices, currently deployable \CFA leverages Trusted Execution Environments (TEEs) along with binary instrumentation.

Unfortunately, existing TEE-based \CFA techniques are not able to guarantee that \CFlog is received by \vrf.  
Although an absence of responses containing \CFlog leads \vrf to distrust \prv (since an honest \prv would respond), it does not support remotely auditing the compromising behavior on \prv.
Once compromised, \prv may refuse to send \CFlog to \vrf, preventing its analysis to determine the root cause of the compromise and appropriate remediation.
\textit{ACFA}~\cite{acfa} has recently acknowledged this problem and proposed hardware modifications to existing MCU architectures to ensure the reliable delivery of runtime evidence to \vrf and to facilitate \vrf-triggered remediation.
However, because ACFA relies on custom hardware extensions, its guarantees cannot be realized until new MCU chips are fabricated. Consequently, no existing technique is directly deployable in today's commercial ``off-the-shelf'' MCUs.
Our work aims to resolve this conflict by making the following contributions:

\begin{myitemize}
    \item  We propose \acro, the first design realizing {\it secure runtime auditing} on off-the-shelf MCUs. \acro's TEE-based approach combines a \CFA Engine and a Supervisor, both of which are implemented within TrustZone's Secure World. The former records control flow transfers to \CFlog, and the latter actively takes over \prv's execution to enforce reliable delivery of \CFlog to \vrf.
    Additionally, \acro supports \vrf-configured remediation if/when compromises are detected. As a consequence, our work demonstrates that runtime auditing and guaranteed remediation are achievable on commodity MCUs, featuring the TrustZone-M TEE.
    As \acro proposes 
    a software framework leveraging TrustZone-M, it requires a clean-slate design without overlapping architectural features with the prior work~\cite{acfa}, while achieving equivalent security guarantees.
    \item We implement and evaluate a fully functional and open-source prototype of \acro (available at~\cite{tracesrepo}) using the well-known ARM Cortex-M33 MCU. \acro prototype realizes runtime auditing/guaranteed remediation and is accompanied by sample use cases targeting on-demand sensing/actuation settings. We also conduct a systematic security analysis, performance evaluation based on several embedded programs, and an empirical evaluation of \acro under exemplary exploits.
\end{myitemize}


\section{Background}  
\label{sec:background}

\subsection{TrustZone for ARM Cortex-M MCUs}
\label{subsec:background_tz}
TrustZone for ARM Cortex-M (TrustZone-M)~\cite{Armv8_M_TZ_spec} is an architectural security extension available on ARM V8 MCUs. It implements a TEE by isolating hardware and software resources on the MCU between two worlds, namely the ``Secure'' and ``Non-Secure'' Worlds~\cite{ARM-TrustZone}.
TrustZone-M hardware controllers isolate the two worlds so that both code and data in the Secure World are immutable and inaccessible to the Non-Secure World.
The Secure World code can only be called by the Non-Secure World from well-defined entry points, called Non-Secure-Callables (NSC). 
This mechanism enables controlled invocation and trustworthy execution of security-critical code, as well as the safe storage of private data within the Secure World, even when the Non-Secure World code (in this case, untrusted MCU applications) is fully compromised.

TrustZone-M defines the security state of memory segments by configuring hardware controllers called the Secure Attribution Unit (\SAU) and the Implementation-Defined Attribution Unit (IDAU) to enforce memory isolation between worlds~\cite{pinto2019demystifying}.
IDAU is a fixed memory map defined by the manufacturer and it assigns a default minimal security level to a given set of addresses. \SAU, on the other hand, can be configured by the Secure World code to further reserve additional parts of the address space to the Secure World.

Memory accesses are first checked according to the security attribution defined by \SAU, then are checked by the Memory Protection Unit (MPU). With TrustZone security extension, the MPU is segmented into Non-Secure and Secure states, effectively establishing one MPU for each world. The Secure MPU registers are only accessible to the Secure World, whereas the Non-Secure MPU (NS-MPU) registers are accessible to both worlds. MPU configuration registers are by default only accessible to software executing in privileged mode~\cite{arm-mpu-spec} and write access to MPU configuration registers can be revoked via the System Configuration Registers~\cite{stm32l552xx}.

TrustZone-M also supports the assignment of interrupts using separate Interrupt Vector Tables (IVTs) for the Secure and Non-Secure Worlds. IVTs are managed by the Nested Vector Interrupt Controller (NVIC). Each interrupt can be assigned as Secure or Non-Secure by setting the Interrupt Target Non-Secure (NVIC\_ITNS) register~\cite{sbis}, which is only configurable by the Secure World code.
All secure interrupts have higher or equal priority than non-secure interrupts.
When a secure interrupt is triggered while the CPU is executing in the Non-Secure World, the CPU pauses its execution, fetches the address stored in the Secure IVT, and transfers execution to the Secure World Interrupt Service Routine (ISR) pointed by this address. The context of the interrupted task is pushed onto the Non-Secure stack and popped upon return from the interrupt to the Non-Secure World.


\begin{figure}[t]
    \centering
    \includegraphics[width=.75\columnwidth]{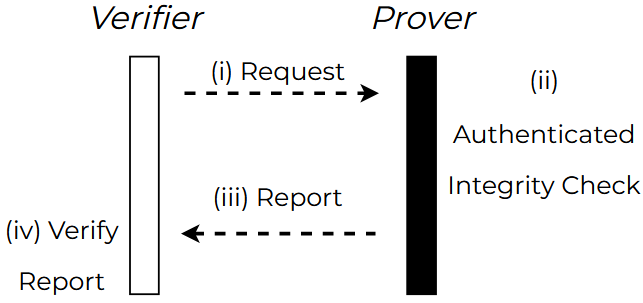}
    \caption{Typical \RA interaction}
    \vspace{-1em}
     \label{fig:RA}
\end{figure}

\subsection{Remote Attestation (\RA)}
\label{subsec:background_ra}

\RA is a challenge-response protocol in which \vrf aims to determine whether a remote \prv is installed with the expected software image.
As depicted in Fig.~\ref{fig:RA}, \RA usually is composed of the following steps:
\begin{myenumerate}
    \item \vrf sends an attestation request to \prv containing a cryptographic challenge $Chal$.
    \item A root of trust on \prv produces a report $H$ by computing an authenticated integrity-ensuring function over \prv's own program memory and $Chal$.
    \item \prv transmits $H$ to \vrf.
    \item \vrf compares $H$ against the expected value to determine if \prv is in a trustworthy state.
\end{myenumerate}

The authenticated integrity-ensuring function in step 2 above can be implemented as a Message Authentication Code (MAC) or a digital signature.
The secret key used in this computation must be securely stored by \prv's root of trust to ensure that it is immutable and inaccessible to any untrusted (potentially compromised) software on \prv.
Therefore, secure storage for the \RA secret key implies some level of hardware support (e.g., from TEEs, as in this work).

\subsection{Control Flow Attestation (\CFA)}
\label{subsec:background_cfa}

\begin{figure}[t]
    \centering
    \includegraphics[width=0.95\columnwidth]{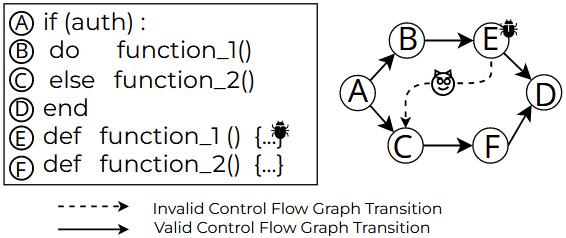}
    \caption{Illustration of a control flow attack}
    \vspace{-0.5em}
     \label{fig:Control FLow attack illustration}
\end{figure}

Control flow attacks~\cite{one1996smashing, shacham2007geometry,ma2023ret2ns}
aim to alter a program's intended control flow by executing unintended sequences of instructions. 
To illustrate these attacks, we refer to Fig.~\ref{fig:Control FLow attack illustration} which shows a simple exemplary program and its control flow graph (\CFG).
In benign executions, there are two valid paths in the program's \CFG: \{$A,B,E,D$\} or \{$A,C,F,D$\} depending on the variable \texttt{auth}.
However, suppose a memory safety vulnerability (e.g., a buffer overflow\cite{war_in_mem}) exists in \texttt{function\_1()} implementation (i.e., node $E$). In that case, it can be exploited to illegally overwrite \texttt{function\_1()}'s return address in the stack modifying the expected return site (node $D$) to an arbitrary address chosen by \adv~-- the call to \texttt{function\_2()} (node $C$) in this example.
As a consequence, the illegal sequence of nodes \{$A,B,E,C,F,D$\} would be executed instead, even though the direct transition from the $E$ to $C$ does not exist in the program's \CFG.
More sophisticated control flow attacks -- such as return-oriented programming (ROP)~\cite{rop} and jump oriented programming (JOP) \cite{jop}) -- can chain multiple such illegal control flow transfers to trigger arbitrary (often Turing-complete) behavior without modifying the program's code. In turn, these attacks cannot be detected by classic \RA protocols.

\CFA~\cite{verif_integrity_mini_survey, tinycfa, oat, dessouky2018litehax, zeitouni2017atrium, scarr, recfa, cflat, geden2019hardware} augments \RA reports to enable detection of control flow attacks. 
In addition to proving whether the correct software image is installed on \prv, \CFA also produces a trace informing \vrf of the order in which the program's instructions have executed.
This trace consists of an authenticated log (\CFlog) of all control flow transfers that occurred during a program's execution.
\CFlog is produced at runtime and stored in protected memory.
Existing \CFA techniques use either (1) binary instrumentation along with TEE support; or (2) custom hardware modifications to generate \CFlog by detecting and saving each branch destination to hardware-protected memory. Upon receiving \CFlog, \vrf can inspect it alongside the attested software image to detect control flow attacks. For instance, in the attack example of Fig.~\ref{fig:Control FLow attack illustration}, the illegal sequence \{$A,B,E,C,F,D$\} would appear on \CFlog and therefore \vrf would distrust this malicious execution.

In the case of TEE-based \CFA (i.e., the class of \CFA approaches applicable to existing devices without requiring custom hardware modifications) TrustZone's Secure World is used as a root of trust to build and store \CFlog. 
The binary to be attested is instrumented so that all branch instructions (e.g., \texttt{jumps}, \texttt{returns}, \texttt{calls}, etc.) are prepended with additional TrustZone calls that trap execution onto the Secure World. Once in the Secure World, \CFlog is updated to reflect the correspondent control flow transfer.

Once execution completes, \prv authenticates \CFlog and the installed software image (as in typical \RA) to produce the attestation response (e.g., by computing a MAC or signature using the attestation secret key).
Finally, \prv transmits \CFlog to \vrf along with the produced authentication token (i.e., the MAC/signature result). In possession of the attested binary, \CFlog, and their authenticator, \vrf can determine if \prv execution occurred as intended and detect control flow attacks (in addition to binary modifications).

\section{\acroplain Overview}
\label{sec:overview}

As noted earlier, \CFA cannot guarantee that \CFlog is received by \vrf when \prv is compromised. 
While this suffices to detect compromises (in general, the absence of a response indicates that \underline{something is wrong}), it does not enable {\it auditing} of \CFlog to pinpoint the source of compromises (i.e., to determine \underline{what is wrong}).
The latter property is non-trivial to obtain since a compromised \prv might ignore the protocol and refuse to send back attestation responses indicating a compromise (and its root cause).

\acro is designed to guarantee that \vrf eventually receives runtime reports containing the information about \prv execution. \acro also supports guaranteed healing of \prv when a compromise is detected. In contrast with prior related work~\cite{acfa}, \acro does not require custom hardware modifications to the MCU, enabling control flow auditing in existing ({\it ``off-the-shelf''}) devices. 

\acro targets auditing on-demand sensing/actuation operations denoted \app-s, where \vrf requests the execution of one particular \app on \prv, at a given time.
It extends the traditional ``\texttt{\vrf sends request $\rightarrow$ \prv executes requested operation $\rightarrow$ \vrf receives result}'' paradigm to enable control flow auditing  (and potential remediation) of each requested operation.
\acro implements a Secure World-resident software monitor. 
Any attested \app is untrusted and thus executes in the Non-Secure World.

\subsection{Goals}

\textbf{Runtime Auditing:} 
\acro guarantees that a compromised \app on \prv cannot interfere with the generation or transmission of a runtime report.
It also ensures that each report is reliably received by \vrf, periodically re-transmitting a report until a subsequent confirmation message is received from \vrf.
In line with prior TEE-based \CFA, \acro instruments \app binary to construct \CFlog by logging all non-deterministic control flow transfers during \app execution.
Each runtime report sent to \vrf contains an authenticated \CFlog of \app's execution (or a partial slice of \CFlog, if the memory region reserved to store \CFlog fills up during \app's execution).
By inspecting the received \CFlog, \vrf can audit \prv's runtime execution and pinpoint the exploit source when an attack is detected.

\textbf{Guaranteed Remediation:}
\acro Secure World implementation retains control over \prv execution after sending a report and until a response is received by \vrf.
If the response indicates that no exploit was found, the Non-Secure World execution simply resumes. 
If \vrf responds with a remediation request, \acro executes a Secure World-resident remediation function immediately. 
Examples include: wiping all data memory, shutting down \prv, or updating its software (i.e., its Non-Secure World program memory section containing the vulnerable \app). 
Remediation actions are configurable by \vrf according to a desired security policy.

\subsection{Architecture at a High-Level} 

\begin{figure}[t]
    \centering
    \includegraphics[width=0.99\columnwidth]{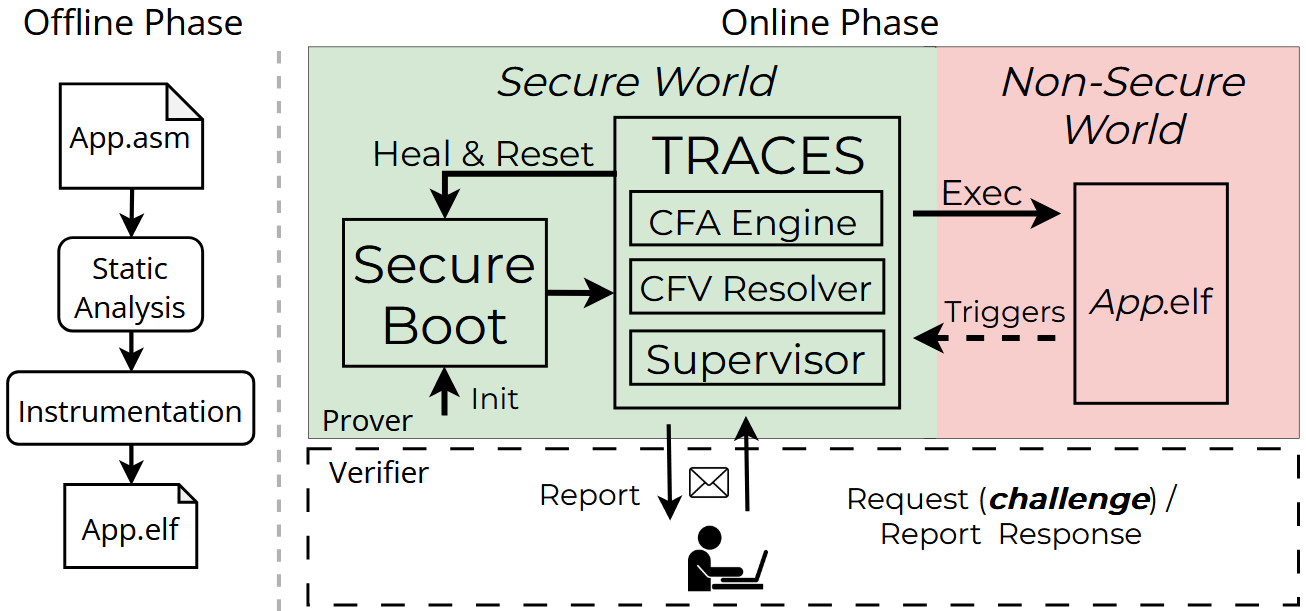}
    \caption{High level illustration of \acro design.}
    \vspace{-1em}
    \label{fig: model overview 1}
\end{figure}

As illustrated in Fig.~\ref{fig: model overview 1}, \acro consists of three Secure World-resident software modules: {\bf (i)} the \CFA engine, {\bf (ii)} the Control Flow Violation (CFV) Resolver, and {\bf (iii)} the Supervisor. \CFA engine is responsible for maintaining \CFlog. CFV Resolver implements \vrf desired remediation action. Lastly, the Supervisor acts as a controller within \acro. It handles the transitions between different \acro actions and enforces the required security properties while executing each of these actions.

Before deployment, \app's binary is instrumented with a call to \CFA Engine at each non-deterministic branching instruction (note that the presence of expected instrumentation is conveyed to \vrf as part of \acro responses -- see below). At boot, \acro workflow enforces that the Supervisor is always the first software to run on \dev.
It performs boot-time configurations to restrict the Non-Secure World's access to security-critical resources and waits for an initial \vrf authorization to initialize the Non-Secure World execution.

At runtime, upon receiving a \vrf-issued request for the attested execution of some \app, the Supervisor measures \app's binary (by hashing the Non-Secure World's program memory), configures the measured region as read-only (to prevent code modifications after the initial measurement), disables Non-Secure World interrupts, and initiates \app's execution 
in the Non-Secure World. During \app execution, \CFlog is continually appended with control flow transfers due to the instrumented \CFA Engine calls.
Aside from \CFA Engine calls, after \app execution is initiated, \acro Secure World implementation acts upon three events,
referred to as triggers [T1], [T2], and [T3] defined as:
\begin{myenumerate}[label=\textbf{[T\arabic*]}:]
\item A predefined time limit for periodic communication with \vrf has been reached. This guarantee is obtained by leveraging the NVIC controller (recall Sec.~\ref{sec:background}) to assign a timer-based interrupt to the Secure World. 
\item The execution of \app has concluded.
[T2] is obtained with a TrustZone-protected return instruction to the Secure World.
\item The memory region reserved to store \CFlog is full and the current \CFlog values need to be sent to \vrf before new control flow transfers can be added. To obtain [T3], \acro checks if \CFlog designated memory is full after appending each new control flow transfer.
\end{myenumerate}

[T1] is implemented as a Secure World-protected interrupt. [T2] is a secure (i.e., TrustZone-protected) return to Secure World that is measured by \app's hash and immutable after that stage. [T3] is implemented within the Secure World code.
Therefore, given TrustZone guarantees, these triggers cannot be disabled by the Non-Secure World.
Furthermore, the Secure World execution, once triggered, cannot be interrupted by the Non-Secure World.
Any of the three triggers results in a call to \acro to generate a signed runtime report containing the current snapshot of \CFlog as well as the hash of \app's binary. The Supervisor sends the report to \vrf for analysis and retains control over \prv in the Secure World until an authenticated acknowledgment response is received from \vrf.

\textit{\textbf{Remark:} note that the hash of \app's binary is sent along with every report, allowing \vrf to also ascertain \app's binary integrity. This implies the presence of the expected code instrumentation responsible for generating \CFlog. See protocol details in Sec.~\ref{sec:details}.}

While waiting for \vrf's response, the produced report is re-transmitted periodically to cope with eventual network losses.
If \vrf response indicates that a compromise was detected, Supervisor invokes CFV Resolver for remediation.
If no compromise is indicated, Supervisor simply resumes \app execution from where it left off when the trigger occurred.
This process continues while \app executes (i.e., until \vrf receives a report issued due to [T2]).

\begin{figure*}[th!]
    \centering
        \includegraphics[width=0.9\textwidth]{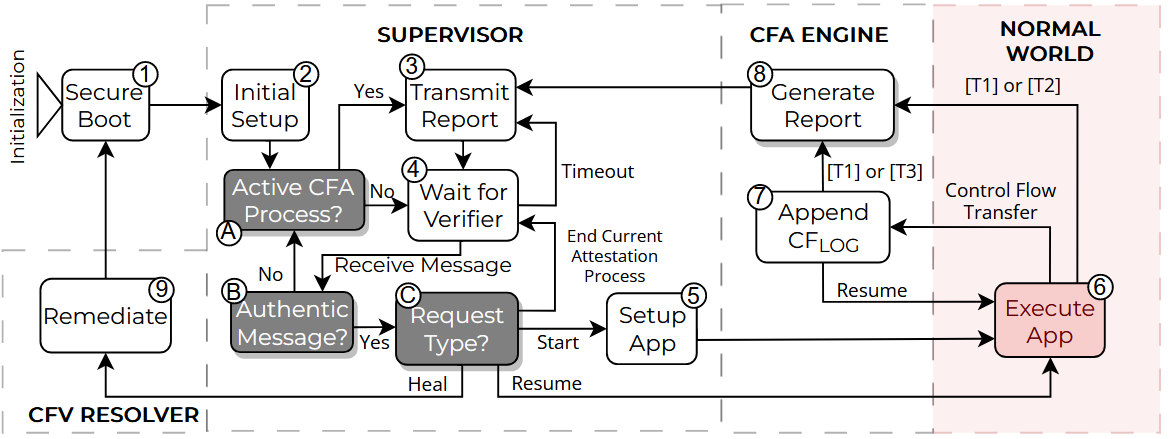}
    \caption{\acro workflow.} 
    \vspace{-1em}
    \label{fig: state model}
\end{figure*}

\subsection{TRACES Security Intuition.}
The intuition for \acro's security (in achieving auditing and guaranteed remediation capabilities) follows from the facts that: {\bf (1)} triggers [T1], [T2], and [T3] cannot be disabled by the Non-Secure World and {\bf (2)} the Supervisor-enforced workflow cannot be disrupted by the Non-Secure World. The workflow assures that \vrf always receives reports and can act upon them before execution of untrusted software in \prv is resumed. In Sec. \ref{sec:details}, we delve into the details of how these high-level ideas are obtained concretely. In Sec.~\ref{sec:sec_analysis}, we perform a systematic security analysis of \acro.

\section{\acroplain in Detail}
\label{sec:details}

\subsection{Scope and System Model}
\label{subsec:sysmodel}

\acro is focused on single-core, bare-metal MCUs equipped with TEEs, (TrustZone-M, in our case). Attested \app-s execute in the Non-Secure World, while the Secure World contains \acro trusted computing base (TCB).
We rely on the following standard requirements from \prv's TEE (which are attainable through standard ARM TrustZone-M v8 architectural support~\cite{Armv8_M_TZ_spec}):

\begin{myitemize}
\item  Cryptographic keys are securely provisioned to \prv and \vrf prior to deployment. \vrf is trusted and \prv keys are stored within the Secure World (thus inaccessible to the Non-Secure World).
\item  The Secure World (code and data) is trusted and isolated from Non-Secure World (including \app);
   
\item  \prv has separate IVTs for the Non-Secure and Secure Worlds. Interrupt sources can be assigned to either world and the Secure-IVT has priority over the Non-Secure-IVT. 
   
\item \prv has a Non-Secure Memory Protection Unit (NS-MPU) that controls access to Non-Secure World memory. 
(see Sec.~\ref{subsec:background_tz}). 
\end{myitemize}

Following the on-demand sensing/actuation regime discussed in Sec.~\ref{sec:overview}, \vrf aims to audit the runtime behavior of one Non-Secure World \app at a time. Nonetheless, \prv may have any number of \app-s installed in its Non-Secure World program memory.
Similar to any TEE-based \CFA, the branching instructions of \app must be correctly instrumented with additional instructions to log the control flow transfers.
This instrumentation is performed at compile-time before \app's deployment on \prv (which can also happen remotely).
In addition, \vrf keeps a copy of the instrumented \app binary (or hash thereof) to verify the received hash of \app's binary included in \prv's response, as described in Sec.~\ref{subsec:background_ra}.

\subsection{Adversary Model}
\label{subsec:advmodel}

Our \adv model is consistent with that of secure systems built atop TEEs. We consider \adv capable of fully compromising the Non-Secure World on \prv. 
\adv can exploit vulnerabilities to launch code-injection attacks, hijack \app's control flow, or perform code-reuse attacks.
In addition, \adv can manipulate Non-Secure World interrupts and their ISRs.
\adv cannot tamper with code or data in the Secure World or circumvent access controls enforced by the TEE hardware. 

\adv's ability to modify Non-Secure World code must be accounted for in \acro's design because \adv could use it to remove \CFA-related instrumentation. \acro leverages TEE controls along with temporally consistent code integrity measurements to ensure that any such attempt is detected by \vrf (see Section~\ref{subsec:states}).
This is in contrast to prior work~\cite{cflat,oat,wang2023ari} that rules out code modifications from their threat model by assumption (the latter also implies an inability to perform benign software updates at runtime~\cite{rata}). Our design shows the latter requirement/limitation as unnecessary.

Invasive physical attacks that modify hardware are out-of-scope of this work, as they require an orthogonal set of physical security measures. For details see~\cite{ravi2004tamper,obermaier2018past}.

\subsection{\acro Workflow}
\label{subsec:states}

\acro workflow is presented in Fig.~\ref{fig: state model}
and this section details each step, including processes (\textbf{\circled{1}{white}{black}} - \textbf{\circled{9}{white}{black}}) and decisions (\textbf{\circled{A}{white}{black}} - \textbf{\circled{C}{white}{black}}), along with details of how the workflow rules are enforced by \acro design at each stage.

\textbf{Initialization Routine:} 
The first step in the Initialization Routine is a standard \textit{\circled{1}{white}{black} Secure Boot} verification that ensures the integrity of the trusted software loaded onto the Secure World, i.e., \acro's implementation.


\acro Supervisor module is the first to execute after secure boot. It performs the \textit{\circled{2}{white}{black} Initial Setup} by reserving a memory region $SW_{MEM^{'}}$, located within the Secure World, to store the runtime report.
To complete initialization, Supervisor retrieves the runtime auditing context. This includes Secure World-resident memory regions that store \CFlog, \CFlog size, and other metadata related to the runtime auditing context.
We note that in most cases, the context is empty at this stage. However, in some instances, it may contain values from a previous malicious execution that led the MCU to a reset (see below).

\textbf{Waiting for \app's Attested Execution Request:}
After completing all initialization tasks, the Supervisor determines if
an active attestation process of some \app was ongoing before boot \circled{A}{white}{black}.
For instance, if a software fault occurred and led to a system reset during \app execution, a report transmission \circled{3}{white}{black}  must occur next to notify \vrf of the runtime state that has led to the fault.
%
If there is no previous operational context (\prv was inactive), Supervisor continues to \textit{\circled{4}{white}{black} Waiting for Vrf Message}. Upon receiving a message from \vrf, the Supervisor checks whether this message is timely and generated by \vrf \circled{B}{white}{black} (via standard cryptographic authentication -- see Sec.~\ref{subsec:protocol} for protocol details).
Upon successful authentication, the Supervisor processes \vrf request \circled{C}{white}{black}. A request to start the attested execution of some \app on \prv contains a unique challenge $\chal$ used to ensure the freshness of the report to be generated by the \app's attested execution.
The Supervisor saves $\chal$ and associated metadata within the Secure World before executing \app.

\textbf{Attested Execution:}
A new attested execution requires the Supervisor to \textit{\circled{5}{white}{black} Setup App} and perform the following actions before beginning execution.
%
\begin{myitemize}
\item Setting the Non-Secure World program memory as Read/Execute-Only and its data memory as Non-Executable using the NS-MPU. This prevents unauthorized code modification and data execution at runtime.

\item Revoking the Non-Secure World's ability to reconfigure NS-MPU. This is achieved in two steps: (1) by setting the System Configuration Registers to prevent accesses to the NS-MPU configuration registers; and (2) by using \SAU to assign the System Configuration registers themselves to be accessible only by the Secure World. After this stage, both the NS-MPU and System Control registers are inaccessible to any Non-Secure World code.

\item Configure [T1] as a secure timer interrupt to activate after $\delta$ clock cycles, with a higher priority than all other interrupts. As a Secure World interrupt, [T1] cannot be disabled or misconfigured by untrusted code in the Non-Secure World. This timer resets every time a report is generated and is deactivated at the end of \app's attested execution. The value of $\delta$ can be either fixed to a default or chosen by \vrf within the attested execution request.

\item Measure \app's code by hashing the entire Non-Secure World's program memory to capture its state immediately before execution. The hash result ($H_{PMEM}$) is stored in Secure World memory. 

\item Initialize \CFA engine metadata. This includes setting the \CFA status flag as ``active''.

\end{myitemize}

Once all configurations are set, the Supervisor \textit{\circled{6}{white}{black} Executes App} in the Non-Secure World. 
During \app's execution, \CFA Engine is invoked by \app instrumented instructions whenever a new control flow transfer occurs to \textit{\circled{7}{white}{black} Append \CFlog}. Each instrumented instruction appends the current control flow transfer to \CFlog and then resumes \app execution \circled{6}{white}{black} as long as \CFlog is not full. If \CFlog is full, \CFA Engine triggers [T3] 
to transmit a new report to \vrf \circled{8}{white}{black} before additional transfers can be added.


\textbf{Reliable Runtime Report \& \vrf Response:}
When either trigger [T1], [T2], or [T3] is activated, \CFA Engine must \textit{\circled{8}{white}{black} Generate a Report} to be sent to \vrf. To ensure atomic execution, all interrupts are disabled as the first step of \circled{8}{white}{black}. Moreover, the timer for [T1] is cleared and paused until \app execution is resumed. The report is cryptographically authenticated (see protocol details in Sec.~\ref{subsec:protocol}) and includes \chal, $H_{PMEM}$, \CFlog, its size, and outputs (e.g., sensed values) produced by \app execution (if any).

After computing the report, Supervisor \textit{\circled{3}{white}{black} Transmits the Report} and waits for a response \circled{4}{white}{black}. While \prv waits, \vrf is expected to: receive the report and verify its authenticity, validate \CFlog and $H_{PMEM}$, and respond to \prv based on this analysis. During this waiting period in \circled{4}{white}{black}, the Supervisor enters sleep mode and periodically wakes up to re-transmit the report in \circled{3}{white}{black}. This periodic re-transmission is necessary to ensure that the report eventually reaches \vrf, despite occasional network failures or network denial of service attempts.
Upon receiving a response from \vrf, \prv proceeds to perform checks \circled{B}{white}{black} and \circled{C}{white}{black}.

If a violation is detected in the report, \vrf message includes a {\it Heal} request (in \circled{C}{white}{black}), causing Supervisor to invoke the remediation routine managed by CFV Resolver in \circled{9}{white}{black}. In the absence of control flow violations, there are two cases. If the previous report was generated by [T2], indicating that \app has completed its execution, \vrf response instructs the Supervisor to stop the current attestation process and return to the idle waiting stage \circled{4}{white}{black}. To do so, Supervisor changes the \CFA status flag to ``inactive'' and waits for a new attested execution request from \vrf. Otherwise, \app's attested execution is resumed \circled{6}{white}{black}.

\textbf{CFV Resolver:}
When \vrf message contains a request to {\it Heal} \prv, the CFV Resolver is invoked following \circled{C}{white}{black} to execute the \textit{\circled{9}{white}{black} Remediation Procedure}. This module contains the \vrf-defined remediation action that can accommodate a variety of policies. For instance, erasing all data memory, updating \app's binary to patch the vulnerability, or shutting \prv down. By default, this action is followed by a system reset to ensure that the system reboots in the newly configured state properly. All non-maskable interrupts and hardware fault exceptions that could otherwise preempt remediation actions are configured to reset \prv, thereby triggering the Supervisor execution at reboot and resuming the remediation action in \circled{9}{white}{black}.

As mentioned earlier, if \prv resets \circled{1}{white}{black}  during \app's attested execution (e.g., due to faults or software bugs) the remnants of \CFlog must be sent to \vrf for auditing (e.g., to pinpoint the exploit causing the reboot). In this case, the \CFA process remains active when Supervisor reaches \circled{A}{white}{black}. At this point, Supervisor continues to \circled{3}{white}{black} instead of \circled{4}{white}{black} and re-transmits the report containing the remnants of \CFlog.


\subsection{ \acroplain Protocol}
\label{subsec:protocol} 

Based on the workflow defined in Sec.~\ref{subsec:states}, we specify the interaction between \vrf and \prv in Protocol 1.
This protocol starts when \prv receives a request from \vrf.

In Step 1, the Supervisor hashes and locks (prevents writes to) the Non-Secure World program memory ($PMEM$) to produce $H_{PMEM}$. 
In Step 2, \acro starts executing \app. \CFlog is appended whenever a control flow transfer happens within \app.
Upon any of the triggers [T1], [T2], or [T3], \prv proceeds to Step 3.

In Step 3, \acro attests \prv state by computing a MAC ($\sigma_\prv$) on $H_{PMEM}$ and \CFlog using a pre-shared symmetric key (an asymmetric version of the protocol can be obtained in the standard way, by replacing the MAC operations by signatures). The runtime report ($R_P$), which includes $\sigma_\prv$, \CFlog and $Log_{Size}$, is then created and sent to \vrf in Step 4.
Upon receiving $R_P$ in Step 5, \vrf proceeds to:
\begin{myenumerate}
\item validate $\sigma_\prv$
by checking whether it was computed over the latest challenge \chal, the
hash of the expected binary $PMEM'$, and 
the received \CFlog.
\item analyze \app execution using the received \CFlog. 
\vrf can complete this step using several techniques, e.g., determining whether \CFlog matches a valid path in \app's CFG, emulating a shadow stack of \app's execution, and more. Sec.~\ref{subsec:verification} elaborates on verification possibilities.
\end{myenumerate}

Based on this analysis, in Step 6, \vrf produces an output ($vrf_{result}$) to indicate the verification result and next action. 
In Step 7, \vrf issues a fresh challenge $\chal'$ and creates an authentication token ($\sigma_\vrf$) by MAC-ing $\chal'$ and $vrf_{result}$ and transmits their response $R_V$ to \prv in Step 8.

In Step 9, \prv receives and parses $R_V$. 
Next, in Step 10, \prv verifies $R_V$ by checking whether the MAC $\sigma_\vrf$ is valid and if $\sigma_\vrf$ was computed on a fresh challenge. 
This step produces a verification output $out$; when $out=\texttt{False}$, \prv disregards the response, continues to wait, and repeats Step 4 until it receives an authenticated message from a valid \vrf.

Otherwise, \prv updates its own persistent copy of the latest challenge to the newly received $\chal'$ and examines $vrf_{result}$ to decide on the next course of action. If \vrf disapproves the report ($vrf_{result} = \texttt{Heal}$), \prv invokes the CFV Resolver to execute the remediation (Step 11) and subsequently restarts the system. Conversely, when \vrf approves the report ($vrf_{result} = \texttt{Exec}$ or $\texttt{End}$), \prv transfers control back to the Non-Secure World to resume or end \app's execution.


\begin{figure}[t]
\footnotesize

\begin{protocol}{\textbf{- \acro Protocol}} 
\vspace{0.25em}
\textit{NOTATION}:

\begin{myitemize}
    \item $PMEM$: \prv's Non-Secure World program memory.
    \item $PMEM'$: Expected \prv's Non-Secure World program memory.
    \item $Log_{size}$: Size of \CFlog. 
    \item $\mathtt{h}$: A secure cryptographic hash function.
    \item $\mathtt{MAC}_{K}$: Compute MAC using key $K$.
    \item $\mathtt{Verify}_{K}$: MAC verification using public key $K$.
    \item $k$: key pre-shared between \prv and \vrf
    \item $\chal$: Challenge based on a (persistent) increasing counter. 
\end{myitemize}
\sbline
\textit{PROTOCOL}:

\textbf{\textit{Prover (\prv) Secure World}}
    \begin{enumerate}[label=\arabic*., align=left, leftmargin=*, nosep]
        \item Generate hash of and lock $PMEM$ (executed before \app's execution.)
        \AtBeginEnvironment{equation*}{\setlength{\abovedisplayskip}{3pt}\setlength{\belowdisplayskip}{3pt}}
        \begin{equation*}
            H_{PMEM} := \mathtt{h}(PMEM)
        \end{equation*}
        \item  Execute \app in the Non-Secure World. During \app's execution, the Secure World is invoked to append \CFlog whenever a control flow transfer happens.
        \item Upon a trigger, compute MAC:

        \begin{equation*}
            \sigma_\prv := \mathtt{MAC}_{k}(H_{PMEM}, Log_{size}, \CFlog, \chal)
        \end{equation*}
        
        \item Send report $R_P$ to \vrf with the following format:
        \begin{equation*}
            R_P:= (\sigma_\prv || Log_{size} || \CFlog) 
        \end{equation*}
         Wait for \vrf's response and re-transmit $R_P$ periodically until the response is received. 
    \end{enumerate}\\

\textbf{\textit{Verifier (\vrf)}}
    \begin{enumerate}[label=\arabic*., align=left, leftmargin=*, nosep]
        \setcounter{enumi}{4}
        \item  Receive $R_P$ and extract $\sigma_\prv, Log_{size} ,\CFlog$ 
        \item Verify report :
        \begin{equation*}
            vrf_{result} := \mathtt{Verify}_{k}(\sigma_\prv,PMEM',Log_{size}, \CFlog, \chal)
        \end{equation*}
        \item Increment $\chal' := \chal+1$ and creates an authorization token based on this new challenge:
        \begin{equation*}
        \begin{aligned}
            \sigma_\vrf  & := \mathtt{MAC}_{k}(\chal',vrf_{result})
        \end{aligned}            
        \end{equation*}
        \item Construct and send response $R_V$ to \prv
        \begin{equation*}
              R_V := (vrf_{result}||\chal'||\sigma_\vrf) \to \prv
        \end{equation*}
    \end{enumerate}\\
    
\textbf{\textit{Prover (\prv) Secure World}}
\begin{enumerate}[label=\arabic*., align=left, leftmargin=*, nosep]
    \setcounter{enumi}{8}
    \item Receive $R_V$ and extract $vrf_{result}, \chal', \sigma_\vrf \gets R_V$
    \item Authenticate the response, producing a one-bit output:
    \begin{equation*}
        out := \mathtt{Verify}_{k}(vrf_{result}, \chal', \sigma_\vrf) \texttt{ and }  (\chal' > \chal)
    \end{equation*}
    Based on $out$ and $vrf_{result}$, it decides the next transition:
    \begin{myitemize}
        \item If $out=\texttt{False}$: Re-enter Wait (Jump to Step 4)
        \item Else If $vrf_{result}=\texttt{Heal}$: Update local value of $\chal$ to $\chal'$ and enter Remediate state (Jump to Step 11)
        \item Else If $vrf_{result}=\texttt{Exec}$, update local value of $\chal$ to $\chal'$ and resume \app (jump to Step 2)
        \item Else If $vrf_{result}=\texttt{End}$, end 
        \CFA process unlocking PMEM and concluding the protocol instance.
    \end{myitemize}
    \item Execute remediation software and restart the system.
\end{enumerate}
\vspace{-.5em}
\label{protocol:comms}
\end{protocol}
\vspace{-5mm}
\vspace{1em}
\hrulefill
\end{figure}



\section{Security Analysis}
\label{sec:sec_analysis}

Our \adv model (see Sec.~\ref{subsec:advmodel}) considers that \adv has full control over \prv's Non-Secure World. To circumvent \acro guarantees \adv must (1) forge a report that is accepted by \vrf and does not correspond to the actual execution of \app; (2) prevent \vrf from receiving a legitimate response (and \CFlog therein); or (3) prevent a \vrf-initiated remediation.

\subsection{Report Forgery}
\label{subsec:resp_forgery}

A runtime report is considered trustworthy if it faithfully reflects the control flow of \app's timely execution as well as its binary. \adv may attempt to manipulate the response message to deceive \vrf
in the following ways:

\textbf{\app Binary Modifications}.
\adv may modify the Non-Secure World binary to remove or add instructions that generate \CFlog entries in an attempt to produce a valid \CFlog that differs from the true control flow of \app's execution. However, \vrf can detect any modifications to \app binary by checking $H_{PMEM}$, which is computed immediately before \app execution. In between $H_{PMEM}$ generation and the end of \app's execution, \app's binary cannot be modified, as enforced using NS-MPU and SAU protections.

\textbf{Control Flow Attacks}.
\adv may attempt to corrupt \app's execution by exploiting memory safety vulnerabilities to cause a malicious sequence of control flow transfers without modifying \app's binary. However, any such attempt must be reflected on \CFlog (due to the instrumentation of all branching instructions) and thus visible to \vrf.

\textbf{Interrupt Manipulation}.
An attacker could also leverage non-secure interrupts to stealthily modify \app's control flow. By default, \acro disables interrupts during \app's execution. For real-time \app-s that must process interrupts, this requirement can also be alleviated by leveraging interrupt-safety mechanisms for \CFA, such as ISC-FLAT~\cite{iscflat}.

\textbf{\CFlog Forgery.}
\adv may attempt to directly forge \CFlog by modifying the response message or the memory region storing \CFlog on \prv. 
Since \CFlog is append-only and stored in the Secure World, \adv cannot modify \CFlog (or other \acro data) in \prv's memory. In addition, attempts to modify or replay a response message are ineffective due to the use of an unforgeable cryptographic function (see below) computed on a fresh challenge (\chal) unique for every response message in the protocol.

\textbf{Forgery of Attestation Result}.
\adv may attempt to forge the attestation result $\sigma_\prv$. 
However, this forgery is computationally infeasible without knowledge of the key given the security of the underlying cryptographic function. Finally, the key is stored in the Secure World, thus inaccessible to \adv.

\subsection{Preventing Evidence Delivery}
\label{subsec:preventing_delivery}

Since triggers [T1], [T2], and [T3] cause \acro to generate and send a report, \adv must prevent them from occurring to block the delivery of evidence to \vrf.
\adv can only avoid filling \CFlog to its maximum size (hence triggering [T3]) by modifying the binary of \app or by launching a control flow attack that jumps to an uninstrumented section of the Non-Secure program memory outside of \app's binary. 
Although these measures prevent [T3], these attacks will always be reflected in the next report caused by triggers [T1] or [T2].
If \CFlog does not fill up to its maximum size during \app execution, \app will eventually end and cause a trigger [T2] or a timeout [T1] (whichever comes first).
Since the timer is configured as a secure interrupt,
it is impossible for \adv to prevent [T1] because it is handled by the Secure World.
The NVIC interrupt configuration is controlled by the Secure World and cannot be modified by the Non-Secure World.

\acro implementation re-transmits evidence until an authenticated confirmation (and remediation request, if applicable) is received from \vrf. 
This guarantees that \vrf does not lose evidence due to network faults/attacks. 
However, it also prevents execution on \prv from resuming before verification is completed successfully, adversely affecting systems that rely on real-time response and time-critical actions.
To cope with this, TRACES can be modified to impose a less-strict policy. This could work by allowing \app to continue execution until the next [T1] trigger, when a new report must be transmitted to \vrf. If no receipt confirmation for the older report is received from \vrf, the new report would now contain both the old \CFlog and new control flow transfers added since \app's resumption. The less-strict verification policy, however, introduces a security trade-off. A compromised \prv could, in this case, execute malicious actions for a longer period, i.e., until the second [T1] trigger. 


\subsection{Preventing Remediation Actions}
\label{subsec:preventing_remediation}

It follows from the atomic execution of the remediation action after communication with \vrf, that \adv cannot prevent it. \acro ensures that \prv execution remains in the Secure World until it has received approval from \vrf to resume the Non-Secure World execution. In addition, \prv stays in the Secure World and attests to the result of any remediation action after its completion. The latter serves as confirmation to \vrf that the remediation action was executed properly. Although \prv may reset, perhaps in an attempt from \adv to prevent the remediation from taking place, 
the report transmission phase is always re-initiated after any reset, and it is eventually followed by \texttt{remediation} (recall Fig.~\ref{fig: state model}). As a result, if a reset indeed occurs, malware in the Non-Secure World is prevented from executing until the remediation phase is completed successfully.


\begin{table*}[t]
\centering
\vspace{-0.5em}
\caption{\app before and after instrumentation}
\vspace{-0.5em}
\label{tab:instrumentation}
\resizebox{.99\textwidth}{!}{%
\begin{tabular}{|c|cccccccc|}
\cline{1-9}
\multirow{2}{*}{Application Information} & \multicolumn{5}{c|}{Sensor Applications} & \multicolumn{3}{c|}{BEEBS Programs~\cite{beebs}} \\
 & Ultrasonic~\cite{ultsensor} & Geiger~\cite{geiger} & Syringe~\cite{opensyringe} & Temperature~\cite{tempsensor} & \multicolumn{1}{c|}{GPS~\cite{gps}} & prime & crc32 & sglib-arraybinsearch  \\ \cline{1-9}
\hline
\rowcolor[HTML]{EFEFEF} Total Instructions & 82 & 223 & 152 & 157 & 1200 & 133 & 57 & 97  \\
Instructions post-instrumentation & 91 & 238 & 172 & 174 & 1389 & 146 & 61 & 116  \\
\rowcolor[HTML]{EFEFEF} Task runtime (ms) & 0.2 & 0.2 & 0.8 & 0.3 & 0.4 & 0.9 & 1.2 & 0.6  \\
Instrumented task runtime (ms) & 0.4 & 1.0 & 1.8 & 1.6 & 3.0 & 5.8 & 9.4 & 13.4  \\
\rowcolor[HTML]{EFEFEF} Control Flow Transfers (at runtime) & 1015 & 744 & 654 & 607 & 649 & 1304 & 2051 & 3225 \\
Generated CFLog size (Bytes) & 56 & 186 & 1400 & 1212 & 2596 & 5216 & 8204 & 12900 \\ \cline{1-9}
\end{tabular}
}
\end{table*}

\section{Implementation Details}
\label{sec:implementation}

\subsection{Instrumentation.}\label{subsec:instr}

Prior to deployment, \app's assembly is instrumented to redirect branches to a trampoline function in the NSC that calls the \CFA Engine. We implement trampoline functions for, indirect calls, conditional branches, and returns. 
    
\textbf{Conditional Branch Instructions.}
Since there are two possible destinations of a conditional branch (``taken" and ``not-taken"), a call to the trampoline function (via \texttt{bl}) is inserted in both places. Then after calling the trampoline, the link register \texttt{lr} holds the branch's destination address. The trampoline passes \texttt{lr} to the \CFA Engine for logging. Afterwards, the \CFA Engine executes \texttt{bxns lr} instruction to return to the branch destination.
%
When a conditional branch occurs due to a static loop (i.e., a loop with no internal branching), it can be instrumented differently to optimize the logging.
A loop is detected in \app's assembly by locating a ``backward" non-linking branch instruction (i.e., a conditional branch instruction whose ``branch-taken" destination precedes the instruction itself)~\cite{zeitouni2017atrium}.
Once located, the register that is incremented (\texttt{ri}) and the register (or value) used as the limit for comparison (\texttt{rL}) are identified.
Then, the instruction that initializes \texttt{ri} is located to determine the loop entry.
At the entry, three instructions are then added.
First, the loop's ``branch-taken" destination is loaded into a reserved register \texttt{rr$_0$}.
Next, the value from \texttt{rL} is loaded into a second reserved register \texttt{rr$_1$}.
Lastly, a \texttt{bl} to a loop trampoline function is inserted. The loop trampoline reads from \texttt{rr$_0$} and \texttt{rr$_1$} to append \CFlog with the destination address and the limit of the loop, respectively.
This optimization significantly reduces the number of calls to \CFA Engine (to just $1$) to log all static loop iterations. 
The loop exit is instrumented like any conditional branch "not-taken" destination.
If the loop is not static or \texttt{rL} is modified in the loop, the conditional branch is instrumented as described earlier, and \CFlog optimization occurs in the \CFA Engine.
In this case, \CFA Engine treats repeated backward edges as loops and increments an internal loop counter instead of logging the repeated address. When the loop exits, the counter is logged.

\textbf{Indirect Call Instructions.}
Indirect calls are of the form \texttt{blx rx}, which calls the address stored in a register \texttt{rx}. To save the value of \texttt{rx}, the trampoline function for indirect calls uses the reserved register \texttt{rr$_0$}. 
Assume an instruction \texttt{blx rx} indirectly calls the function \texttt{func}. This instruction in \app is replaced with two instructions: first, an instruction to load \texttt{rx} into \texttt{rr$_0$}; then, a \texttt{bl} to the trampoline is inserted. When reaching the trampoline, \texttt{rr$_0$} holds \texttt{func}'s address, and \texttt{lr} holds \texttt{func}'s return address. The trampoline then passes the value of \texttt{rr$_0$} to the \CFA engine to update \CFlog. After appending \CFlog, the trampoline returns to \app via \texttt{bxns rr$_0$} and resumes \app at the first instruction of \texttt{func} while preserving \texttt{func}'s return address in \texttt{lr}.
Our implementation reserves \texttt{r10} as \texttt{rr$_0$} and \texttt{r11} as \texttt{rr$_1$}. All other instances of \texttt{r10} and \texttt{r11} in \app assembly are replaced with different general-purpose register before recompilation.

\textbf{Return Instructions.}
There are two scenarios for returns from a function in \app's assembly. If the function performs no other calls, the return is implemented as \texttt{bx lr}. If it contains a call, \texttt{lr} must be pushed onto the stack, and thus the return is implemented as \texttt{pop pc}. 
In the first case, we replace \texttt{bx lr} with a direct branch (via \texttt{b}) to the trampoline. A direct branch ensures the return address in \texttt{lr} is not modified. After logging \texttt{lr}, the trampoline returns via \texttt{bxns lr} to the proper destination. 
In the second case, \texttt{pop pc} is replaced with two instructions: a \texttt{pop lr} followed by a direct branch (via \texttt{b}) to the trampoline. The logging and return are then performed in the same manner as the first scenario.

\subsection{Module Configurations:}\label{subec:impl-details}

In our prototype, the Supervisor communicates with \vrf using a UART-to-USB connection. The available LP-UART interface is configured at \texttt{baud} rate of 921600 bps. The Supervisor also reserves the MCU's Timer \#3 to the Secure World to implement [T1] with a deadline of 5 seconds and sets the maximum \CFlog size to $50$ KBytes. 
\CFA Engine implementation uses SHA256 and HMAC-SHA256 from HACL\text{*}~\cite{hacl} formally verified library for hash and MAC computations. 
We use default parameters with 256-bit keys, and 512-bit \chal.
In our prototype, CFV Resolver implements three simple remediation actions: freeze \prv execution (i.e., run an infinite loop); disable the compromised \app; and wipe the compromised \app from the Non-Secure World.

The memory region $SW_{MEM^{'}}$ that stores the runtime report must be recoverable through software resets, e.g., by assigning $SW_{MEM^{'}}$ to a persistent memory location. Alternatively, some ARM Cortex-M devices with two SRAM segments allow one segment to be retained when the internal voltage regulator powers off as a feature of the lowest-power mode~\cite{stm32l552xx}. In this work, we configure SRAM2 to retain the runtime report through software resets.


\section{Prototype Evaluation}
\label{subsec:evaluation}
We implement and test \acro using a NUCLEO-L552ZE-Q development board shown equipped with an STM32L552ZE MCU 
based on the ARM Cortex-M33 (v8) operating at 110 MHz. The MCU supports ARM TrustZone-M. 
We evaluate \acro usage on a set of open-source sensor applications\footnote{Some applications required small modifications (ports) to run on ARM Cortex-M.} and on programs from the BEEBS benchmark suite~\cite{beebs}, detailed in Table~\ref{tab:instrumentation}. Additionally, we implement all \vrf's operations in Python.
\acro TCB is implemented with 2383 lines of C code. The Supervisor (including the NSC) accounts for 1250 lines, the CFA Engine -- 861 lines (including SHA256 and HMAC formally verified implementations), and the CFV Resolver -- 272 lines. In total, \acro TCB requires $30.8$ KBytes of program memory.



Table~\ref{tab:instrumentation} details the tested applications.
For the evaluated applications, instrumentation alone adds 0.2-12.8ms of runtime overhead.
Due to the lack of custom hardware to detect branches, the same instrumentation is required in any TEE-based \CFA, irrespective of \acro added guarantees.
Repeated loops with internal branching cause more significant increases in programs like \textit{prime}, \textit{crc32}, and \textit{sglib-arraybinsearch} (abbreviated \textit{search}).
%
%
Table~\ref{tab:instrumentation} also shows the number of control flow transfers in each application execution. 
We note, however, that this number does not always reflect the size of \CFlog due to \acro optimization to replace static loops with counters (recall Sec.~\ref{sec:implementation}).

\subsection{End-to-End Runtime}
\label{subsec:runtime_comparison}

We measure the end-to-end runtime to execute these applications, i.e., the time between \vrf's initial request and the receipt of \acro's last report at the end of \app's execution. The end-to-end runtime depends on several factors, such as the characteristics of the application (shown in Table~\ref{tab:instrumentation}) and the configuration of \acro triggers. Among those, the size of \CFlog dedicated memory has a strong influence on the overhead because it determines the rate of [T3] trigger, affecting the time spent generating and transmitting each partial execution report. 

We compare the end-to-end \app runtime in three settings: original \app without generation of any runtime evidence (baseline), \app with best-effort \CFA (ISC-FLAT~\cite{iscflat}), and \app with \acro runtime auditing/guaranteed remediation guarantees.
For a fair comparison to standard \CFA, we set \acro [T1] and [T3] triggers to large values to ensure that only one report is transmitted by \acro. This is because most \CFA approaches do not support the delivery of partial \CFlog-s, simply aborting execution when \CFlog-dedicated memory is full.

\begin{figure}[t]
    \centering
    \includegraphics[width=0.99\columnwidth]{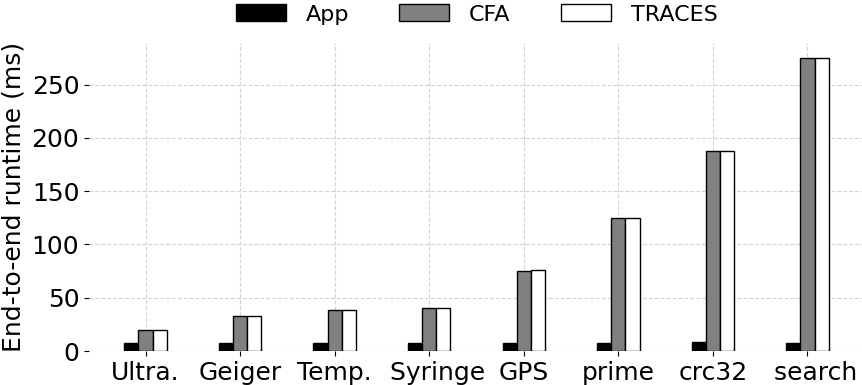}
    \caption{End-to-end runtime comparison: baseline \app, best-effort \CFA, and runtime auditing with \acro.}
    \label{fig:runtime-breakdown}
\end{figure}

Fig.~\ref{fig:runtime-breakdown} presents this comparison. Measured times are the average of 20 executions of each step in each application. Standard deviations in all cases are less than $1\%$ and omitted from the figure.
The end-to-end time varies linearly with the size of \CFlog, resulting in $\approx$12.5-268ms of additional runtime across different applications.

\begin{figure}[t]
    \centering
    \includegraphics[width=0.99\columnwidth]{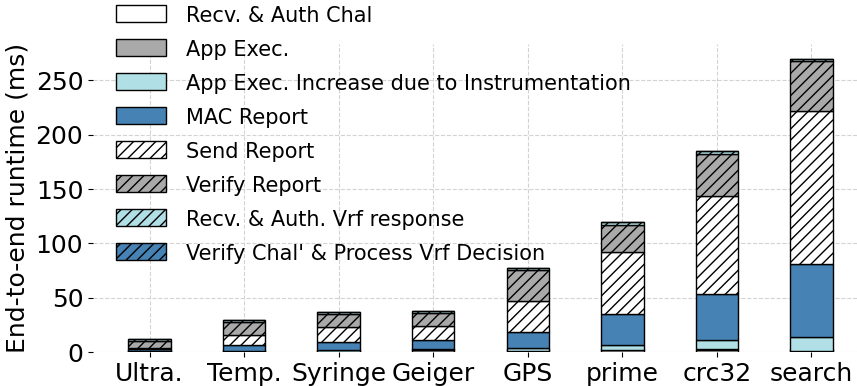}
    \caption{End-to-end runtime breakdown according to steps in Protocol 1. 
    }
    \label{fig:traces-breakdown}
\end{figure}

Fig.~\ref{fig:traces-breakdown} further breaks down \acro end-to-end runtime overheads (for each tested \app) according to individual steps in \acro's protocol. The time associated with receiving and authenticating messages from \vrf is application-independent. On the other hand, the overhead associated with the execution of instrumented instructions, as well as MAC-ing/transmitting/verifying the report, increases with the complexity of \app-s. The only steps unique to \acro, compared to \CFA, are the steps for receiving and authenticating \vrf's response (containing $\chal'$ and their next-action decision) and steps to process this response.

Fig.~\ref{fig:traces-breakdown} shows that the total time taken is dominated by the transmission of generated reports, especially for applications with large \cflog-s. Hence, the size of \cflog-s influences the total end-to-end runtime the most. Similar to Fig.~\ref{fig:runtime-breakdown}, times reported in Fig.~\ref{fig:traces-breakdown} are the average of 20 measurements. Standard deviations in all cases are less than $1\%$.



\begin{figure}[t]
    \centering
    \includegraphics[width=0.99\columnwidth]{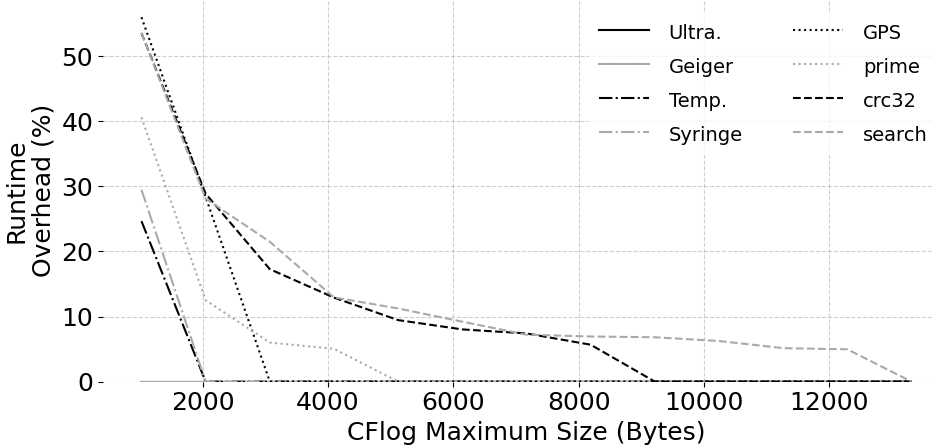}
    \caption{End-to-end overhead as \CFlog increases} 
    \label{fig:end-to-end}
\end{figure}

Next, we experiment by varying the size of \CFlog from 1 to 13 KBytes. In this experiment, we disable the remaining triggers ([T1], [T2]) in order to guarantee that only [T3] triggers occur. Fig.~\ref{fig:end-to-end} shows \acro's end-to-end runtime increase compared to the baseline, where \prv contains an unlimited \CFlog storage and generates one report (i.e., the \acro runtime reported in Fig.~\ref{fig:runtime-breakdown}).

\label{subsec:runtime_breakdown}

\subsection{Attack Detection Delay}
\label{subsec:attack_delay}

\begin{figure}[t]
\centering
    \includegraphics[width=0.95\columnwidth]{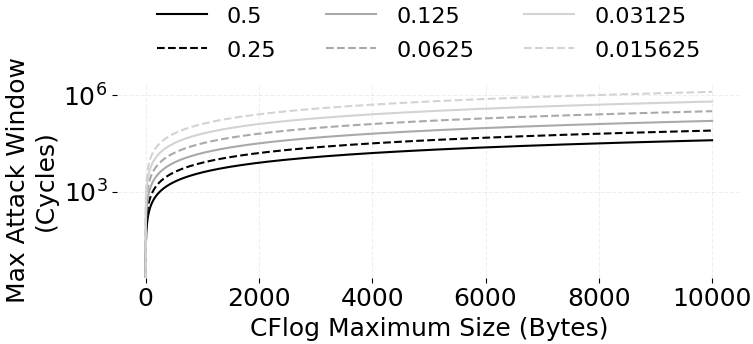}
    \caption{Attack window as \CFlog increases}
    \label{fig:performance}
\end{figure}

While \acro ensures that \CFlog is received by \vrf, there is a brief period (in between triggers) in which detection by \vrf is delayed.
We refer to this period as the maximum ``attack window''.
Two parameters can impact the attack window: (1) the period of [T1], which determines the maximum frequency of report generation, and (2) the maximum size of \CFlog.
The impact of the maximum \CFlog size on the attack window further depends on the ``branch density" of \app, i.e., \app's rate of branch instructions executed. Fig.~\ref{fig:performance} illustrates the relationship between the maximum \CFlog size and the attack window (in CPU cycles) for various branch density values (each line in Fig.~\ref{fig:performance} representing a different density). We note that the time to communicate and verify the report does not affect the attack window. 
This is because during that time, \prv remains in the Secure World.
%
%

If the branch density remains constant, decreasing the maximum \CFlog size will lead to a shorter attack window; increasing \CFlog size will result in a longer attack window. Furthermore, a shorter attack window reduces the time for \adv to interfere with \app's execution. However, it also causes an increase in the end-to-end runtime due to the generation of more reports. This results in an interesting trade-off for \acro parameter choices. For optimal performance in the on-demand sensing setting, [T1] and [T3] should be configured to meet the time and \CFlog storage required for \app's benign execution. Alternatively, [T1] and [T3] can be set to lower values to reduce the attack window by generating more frequent reports (at the price of more frequent transmission overhead).

Finally, we note that the less-strict verification policy discussed in Sec.~\ref{subsec:preventing_delivery} further impacts the attack window
since the time for \vrf to detect the malicious behavior in the less-strict setting could increase by up to one [T1] trigger period.


\subsection{Exploit Detection and \CFA Verification}
\label{subsec:verification}

To exemplify \acro functionality, we evaluate a vulnerable program containing a crafted exploit. We implement a sensor system intentionally designed to contain a buffer overflow vulnerability. In this \app, \prv reads from an input buffer to determine which sensor software to execute (Ultrasonic, Temperature, or both) and how many readings to perform. This application reads from the input until a stop character is received. After reading the input, the command is parsed, and the sensor readings are performed based on the parsed input. Since there is no array bound check while reading from the input buffer, the return address stored on the stack can be overwritten to cause arbitrary behavior; in our example, malicious input causes the program to return to the function that reads the input, effectively causing an infinite loop and preventing any sensor readings from occurring thereafter. 
Since this attack overwrites a return address, the malicious return address value is written to \CFlog.
Based on \prv's runtime report, \vrf must determine if \prv has been compromised. 
\vrf first always checks the authenticity of the report and $H_{PMEM}$ to ensure that \app's instrumented binary was indeed executed. 
If these checks succeed, as they do with the example attack, \vrf examines \CFlog. 
To verify \CFlog, \vrf emulates a shadow stack to validate return addresses shown in \CFlog and utilizes \app's \CFG to validate the execution trace.
As the (significantly more expensive) verification process is outsourced to \vrf, it does not incur additional overhead on the \prv.
After identifying the illegal address caused by the buffer overflow, \vrf initiates remediation.
This example, including the vulnerable software and resulting \CFlog, is available on \acro prototype public repository~\cite{tracesrepo}.


\section{Related Work}\label{sec:rw}
\textbf{Remote Attestation (\RA):}
\RA architectures are generally classified in three types: software-based (or keyless), hardware-based, and hybrid. Software-based architectures~\cite{KeJa03, SPD+04, SLS+05, SLP08, pistis, simple, surminski2021realswatt, scraps} require no hardware support. However, they require strong assumptions about \adv capabilities and implementation optimality.
Hardware-based architectures~\cite{PFM+04, KKW+12, SWP08, sacha, Sancus17} rely on standalone cryptographic coprocessors (e.g., TPMs~\cite{tpm}) or complex support from the CPU instruction set architecture (e.g., Intel SGX~\cite{sgx-explained}). 
For contexts in which standalone hardware is too costly, hybrid architectures~\cite{vrased, smart, tytan, trustlite, hydra, brasser2016remote} have been proposed. Hybrid architectures aim to combine the low hardware cost of software-based approaches with the security guarantees offered by hardware-based approaches through hardware/software co-designs for \RA.
Hybrid \RA has also been extended to provide \vrf with a Proof-of-Execution (\PoX)~\cite{apex, asap} as unforgeable proof that an attested binary executed its outputs were generated by the execution.


\textbf{CFA and Runtime Attestation:}
C-FLAT~\cite{cflat} proposed the concept of \CFA by
leveraging ARM TrustZone to detect and log control flow transfers.
In C-FLAT, the binary is instrumented at branching instructions to trap execution to the Secure World.
Once in the Secure World, the branching information is added to a hash chain. At the end of execution, C-FLAT's hash chain produces a unique value representing the attested program's control flow path. Similarly to C-FLAT, \acro and many later techniques use \app instrumentation and TEE support for \CFA~\cite{scarr, recfa, oat, wang2023ari}.
To remove the requirement of instrumentation, several hardware-based \CFA approaches~\cite{lofat, dessouky2018litehax, zeitouni2017atrium} propose using customized hardware to handle detecting control flow transfers and attesting \CFlog, thus removing the need for instrumentation and requiring a TrustZone-equipped device. 
In hardware-based \CFA, the CPU is extended with dedicated hardware for detecting branch instructions and extracting information about the control flow event (such as the source/destination addresses and type of branch instruction). In addition, they use a hardware-based hash engine to produce a \CFA report for \vrf.
Although these techniques remove the requirement for instrumentation, the hardware cost of a fully hardware-based solution is too costly for MCUs. 
Tiny-CFA~\cite{tinycfa} reduces the hardware cost significantly by executing an instrumented binary atop APEX~\cite{apex}, a low hardware-cost \PoX architecture. Since only minimal hardware changes are required, Tiny-CFA demonstrates \CFA that is realistic for low-end MCUs. 
ACFA~\cite{acfa} reduces the hardware cost without requiring instrumentation by implementing hardware for branch detection/logging and using software for generating the report.

C-FLAT and other early approaches return a single hash of \CFlog, putting \vrf at risk of path explosion during the verification process. To avoid this problem, several recent approaches record a verbatim log of all control flow transfers~\cite{acfa, tinycfa, dialed}. A limitation to logging all control flow transfers verbatim is that \CFlog quickly fills all memory available to the low-end MCU. Because of this, several approaches aim to reduce the size of the verbatim \CFlog by only logging control flow transfers that cannot be determined statically in their entirety~\cite{oat, dessouky2018litehax, recfa, wang2023ari} by reducing the storage size and sending send a series of intermediate log slices~\cite{scarr, acfa}, recording encodings of full addresses~\cite{oat, dessouky2018litehax, recfa}, or recording forward edges verbatim alongside a hash-chain of the return addresses~\cite{oat, wang2023ari}.
\acro logs only non-deterministic branches and supports \CFlog slicing to provide \vrf fine-grained reports while incurring minimal and fixed memory overheads.
%

\begin{table}[t]
    \centering
    \vspace{-0.5em}
    \caption{Qualitative comparison to related work}
    \vspace{-0.5em}
    \label{tab:comp}
    \resizebox{0.99\columnwidth}{!}{%
    \begin{tabular}{|c|c|c|c|c|c|c|c|}
    \hline
      & 'Off-the-shelf' & Verbatim & \CFlog & Runtime & Remote & Runtime & Hardware\\
     Related Work & Support & \CFlog & Slicing & Auditing & Healing & Overhead & Overhead\\
     \hline
     \rowcolor[HTML]{EFEFEF} C-FLAT~\cite{cflat} & Yes & No & No & No & No & Yes & No \\
     OAT~\cite{oat} & Yes & No & No & No & No & Yes & No \\
     \rowcolor[HTML]{EFEFEF} ARI~\cite{wang2023ari} & Yes & No & No & No & No & Yes & No \\
     LO-FAT~\cite{lofat} & No & No & No & No & No & No & Yes \\
     \rowcolor[HTML]{EFEFEF}ATRIUM~\cite{zeitouni2017atrium} & No & No & No & No & No & No & Yes \\
     LiteHAX~\cite{dessouky2018litehax} & No & Yes & Yes & No & No & No & Yes\\
     \rowcolor[HTML]{EFEFEF} Tiny-CFA~\cite{tinycfa} & No & Yes & No & No & No & Yes & Yes\\
     \hline
     \textbf{\acro} & \textbf{\textit{Yes}} & \textbf{\textit{Yes}} & \textbf{\textit{Yes}} & \textbf{\textit{Yes}} & \textbf{\textit{Yes}} & \textbf{\textit{Yes}} & \textit{\textbf{No}} \\
     \hline
    \end{tabular}
    }
    \vspace{-0.5em}
\end{table}

\textbf{Comparison to Related Work.} Table~\ref{tab:comp} compares \acro to most closely related runtime attestation techniques. Similar to our work, C-FLAT~\cite{cflat}, OAT~\cite{oat}, and ARI~\cite{wang2023ari} propose \CFA for off-the-shelf ARM MCUs equipped with TrustZone-M. Thus, they incur a similar runtime overhead due to the required instrumentation for recording control flow events. 
Unlike these techniques, \acro actively triggers the transmission of whole or partial \CFlog-s, guarantees delivery of this evidence, and enables \vrf-triggered device healing in commercial MCUs. 
Tiny-CFA~\cite{tinycfa} combines instrumentation atop a hardware-assisted \PoX architecture~\cite{apex} to achieve \CFA. Hence, it also incurs runtime overhead. Additionally, Tiny-CFA requires custom hardware support That is not available off-the-shelf. LO-FAT~\cite{lofat},
ATRIUM~\cite{zeitouni2017atrium}, and LiteHAX~\cite{dessouky2018litehax} are fully-hardware based approaches. In contrast to \acro, they avoid the runtime overhead cost in exchange for additional hardware features. Because of this, they require new MCUs to be fabricated before they can be used. Similarly, these are techniques for attestation rather than \textit{auditing}, so they do not offer the same guarantees as \acro.

\textbf{Active CFA (ACFA) and Device Healing:}
As discussed in Sec.~\ref{sec:intro}, ACFA is a hybrid (software/hardware) based approach to augment the capabilities of \CFA by leveraging the concept of ``active roots of trust''~\cite{garota} to provide control flow auditing.
As \CFlog is built, a (custom hardware-based) active root of trust interrupts execution to attest the binary and \CFlog.
ACFA also supports a remediation phase after a compromise is detected.
However, contrary to \acro, ACFA requires custom hardware modifications and is therefore not applicable to current devices. 
Other approaches also discuss remote device healing~\cite{healed, pure, ammar2020verify, huber2020lazarus}. Unlike \acro, they either rely on custom hardware, do not consider runtime auditing, and/or use attestation (without reliable delivery) to verify the healing action without guaranteeing its occurrence when \prv is compromised.
%
%


\textbf{ARM TrustZone:}
Prior work has used TrustZone-M to enhance various aspects of embedded system security: availability in real-time systems~\cite{wang2022rt}, low latency secure interrupts~\cite{sbis}, Address-Space Layout Randomization (ALSR) without memory management units~\cite{luo2022faslr}, and virtualization~\cite{pinto2019virtualization}. For a more comprehensive discussion of TrustZone and related applications, we refer the reader to~\cite{pinto2019demystifying}.
\section{Conclusion}
\label{sec:conclusion}

We proposed \acro: an approach for runtime auditing and guaranteed remediation aimed at commodity MCUs equipped with TEEs.
\acro guarantees that control flow logs that contain compromise evidence are always received by a remote verifier. This enables analysis of these logs to pinpoint unknown vulnerabilities and support their remediation. To our knowledge, \acro is the first design to support these security services without requiring custom hardware modifications. We implement a fully functional and open-source \acro prototype~\cite{tracesrepo} based on the ARM TrustZone-M TEE atop the commodity ARM Cortex-M33 MCU.

\section*{Acknowledgements}
We sincerely thank the paper’s anonymous shepherd and ACSAC'24 reviewers for their constructive comments and feedback. RIT authors were partly funded by the National Science Foundation (SaTC award \#2245531).
PSU author was partly supported by the ASEAN IVO (\url{www.nict.go.jp/en/asean_
ivo/}) project, Artificial Intelligence Powered Comprehensive Cyber-Security for Smart Healthcare Systems (AIPOSH),
funded by NICT (\url{www.nict.go.jp/en/}).


\balance
{
\bibliographystyle{IEEEtran}
\bibliography{references}
}
\end{document}